\documentclass[prd,showpacs,amssymb,amsmath,amsfonts,aps,nofootinbib,notitlepage]{revtex4}
\pdfoutput=1 

\usepackage{hyperref}
\usepackage{graphicx}
\usepackage{color}

\DeclareMathAlphabet{\mathsc}{OT1}{cmr}{m}{sc}
\newcommand {\ignore}[1]{}

\def\10{$SO(10)$}
\def\21{SU(2) $\otimes$ U(1) }

\def\422{$SU(4) \otimes SU(2) \otimes SU(2)$}
\def\321{SU(3) $\otimes$ SU(2) $\otimes$ U(1)}
\def\gsim{\raise0.3ex\hbox{$\;>$\kern-0.75em\raise-1.1ex\hbox{$\sim\;$}}}
\def\lsim{\raise0.3ex\hbox{$\;<$\kern-0.75em\raise-1.1ex\hbox{$\sim\;$}}}

\def\lsim{\raise0.3ex\hbox{$\;<$\kern-0.75em\raise-1.1ex\hbox{$\sim\;$}}}
\def\gsim{\raise0.3ex\hbox{$\;>$\kern-0.75em\raise-1.1ex\hbox{$\sim\;$}}}
\def\vev#1{\left\langle #1\right\rangle}
\def \znbb {0\nu\beta\beta}

\newcommand{\AddrAHEP}{%
  AHEP Group, Institut de F\'{\i}sica Corpuscular --
  C.S.I.C./Universitat de Val{\`e}ncia \\
  Edificio Institutos de Paterna, Apt 22085, E--46071 Valencia, Spain}

\newcommand{\ba}{\begin{array}}
\newcommand{\ea}{\end{array}}
\relax
\def\321{$SU(3)\times SU(2)\times U(1)$}

\begin{document}

\begin{flushright}
IFIC/11-03
\end{flushright}

\renewcommand{\Huge}{\Large}
\renewcommand{\LARGE}{\Large}
\renewcommand{\Large}{\large}
\def \znbb {$0\nu\beta\beta$ }
\def \nbb {$\beta\beta_{0\nu}$ }
\title{Phenomenology of Dark Matter from $A_4$ Flavor Symmetry}  
\author{M.~S.~Boucenna}\email{boucenna@ific.uv.es} 
\author{M.~Hirsch}\email{mahirsch@ific.uv.es}  
\author{S.~Morisi}\email{ morisi@ific.uv.es}
\author{E.~Peinado}\email{epeinado@ific.uv.es}
\author{M.~Taoso\footnote{Multidark fellow}}\email{taoso@ific.uv.es}
\author{J.~W.~F.~Valle}\email{valle@ific.uv.es}
\affiliation{\AddrAHEP}

\begin{abstract}
  We investigate a model in which Dark Matter is stabilized by means
  of a $Z_2$ parity that results from the same non-abelian discrete
  flavor symmetry which accounts for the observed pattern of neutrino
  mixing.
  In our $A_4$ example the standard model is extended by three extra
  Higgs doublets and the $Z_2$ parity emerges as a remnant of the
  spontaneous breaking of $A_4$ after electroweak symmetry breaking.
  We perform an analysis of the parameter space of the model
  consistent with electroweak precision tests, collider searches and
  perturbativity.
  We determine the regions compatible with the observed relic dark
  matter density and we present prospects for detection in direct as
  well as indirect Dark Matter search experiments.
\end{abstract}


\maketitle

\section{Introduction}
\label{intro}

The existence of non-baryonic Dark Matter (DM) is well established by
cosmological and astrophysical probes. However, despite the great
experimental effort over many years, its nature still remains
elusive. Elucidating the long-standing puzzle of the nature of dark
matter constitutes one of the most important challenges of modern
cosmology and particle physics.

The various observations and experiments, however, constrain some of
its properties \cite{Bertone:2004pz,Taoso:2007qk}.  Among the most
important requirements a DM candidate is required to satisfy are
neutrality, stability over cosmological time scales, and agreement
with the observed relic density.
While the neutrality of the particle is usually easy to accomodate in
models, its stability in general is assumed in an {\it ad-hoc}
fashion. From a particle physics point of view, the stability
suggests the existence of a symmetry that forbids the couplings that
would otherwise induce the decay. Typically, the most common way to
stabilize the DM particle is to invoke a $Z_2$ parity, an example of
which is R parity in supersymmetry.

It would certainly be more appealing to motivate such a symmetry from
a top-down perspective. Different mechanisms have
been suggested to achieve this \cite{Hambye:2010zb}, for instance using $U(1)$ gauge
symmetries \cite{Frigerio:2009wf,Kadastik:2009dj,Batell:2010bp} to get
for example R-parity in the MSSM from a
$U(1)_{B-L}$\cite{Martin:1992mq}) symmetery, global symmetries,
accidental symmetries \cite{Cirelli:2005uq} or custodial symmetry.

A new mechanism of stabilizing the DM has been recently proposed in
Ref.~\cite{Hirsch:2010ru} in which DM stability originates from the
flavor structure of the standard model.  Indeed the same discrete
flavor symmetry which explains the pattern of neutrino
mixing~\cite{Schwetz:2008er} can also stabilize the dark matter
\footnote{Models based on non-Abelian discrete symmetries but with a
  decaying dark matter candidate can be found for example in
  Ref.\cite{Kajiyama:2010sb}.}.  This opens an attractive link between
neutrino physics and DM~\footnote{Other mechanisms of relating DM and
  neutrinos include the majoron
  DM~\cite{Berezinsky:1993fm,Lattanzi:2007ux,Bazzocchi:2008fh}.}; two
sectors that show a clear need for physics beyond the Standard Model.

The model proposed in Ref.~\cite{Hirsch:2010ru} is based on an $A_4$
symmetry extending the Higgs sector of the SM with three scalar
doublets. After electroweak symmetry breaking two of the scalars of
the model acquire vacuum expectation values (vev) which spontaneously
break $A_4$ leaving a residual $Z_2$.  The lightest $Z_2$ neutral odd
scalar is then automatically stable and will be our DM candidate.

On the other hand, the fermionic sector is extended by four right handed 
neutrinos which are singlets of \321. Light neutrino masses are generated via a
type I see-saw
mechanism~\cite{Minkowski:1977sc,gell-mann:1980vs,yanagida:1979,mohapatra:1980ia,schechter:1980gr},
obey an inverted hierarchy with $m_{\nu3}=0$ and vanishing reactor
neutrino angle \footnote{For a similar realisation see~\cite{Meloni:2010sk}.}.
For pioneer studies on the use of $A_4$ for neutrino physics see
\cite{Babu:2002dz}.

We study the regions in parameter space of the model where the correct
dark matter relic density is reproduced and the constraints from
accelerators are fullfilled.  We then consider the prospects for
direct dark matter detection in underground experiments. We show that
the model can potentially explain the DAMA annual modulation data
\cite{Bernabei:2010mq,Bernabei:2000qi} as well as the excess recently
found in the COGENT experiment \cite{Aalseth:2010vx}.  We show that
present upper limits on the spin independent DM scattering cross
section off nucleons can already severely constrain the parameter space
of the model.
Indirect dark matter searches through astrophysical observations are
not currently probing the model apart from some small regions of the
parameter space where the dark matter annihilation cross section is
enhanced via a Breit-Wigner resonance.

The paper is organized as follows: in Sec.\ref{model} we present the
model, in Sec.~\ref{constraints} the constraints from collider data
are reviewed and in Section~\ref{relic} we study the viable regions of
the parameter space.  In Sec.~\ref{direct} and \ref{indirect} we
sketch the prospects for direct and indirect dark matter detection.
Finally, we summarize our conclusions in Sec.~\ref{conclusions}.

\section{The Model}
\label{model}
 We now provide a concrete realization of dark matter
  based upon the $A_4$ flavor symmetry adopting the simple type-I
  seesaw
  framewok~\cite{Minkowski:1977sc,gell-mann:1980vs,yanagida:1979,mohapatra:1980ia,schechter:1980gr}.
  In order to be phenomenologically viable, the model should account
  not only for the mixing angles describing the observed pattern of
  neutrino oscillations~\cite{Schwetz:2008er} but also for the two
  independent square mass splittings characterizing the ``solar'' and
  ``atmospheric'' sectors.
  These points were taken into account in the model proposed
  in~\cite{Hirsch:2010ru} which we now adopt, and consider its
  phenomenological features in more detail.  The matter fields are
assigned to irreducible representations of the group of even
permutations of four objects $A_4$~\cite{Hirsch:2010ru}, which is
isomorphic to the symmetry group of the tetrahedron, for more details
see Appendix A.  The standard model Higgs doublet $H$ is assigned to a
singlet representation, while the three additional Higgs doublets
$\eta=(\eta_1,\eta_2,\eta_3)$ transform as an $A_4$ triplet, namely
$\eta\sim 3$. The model has in total four Higgs doublets, implying the
existence of four CP even neutral scalars, three physical
pseudoscalars, and three physical charged scalar bosons.

In the fermion sector we have four right-handed neutrinos; three
transforming as an $A_4$ triplet $N_T=(N_1,N_2,N_3)$, and one singlet
$N_4$.  Quarks are assigned trivially as $A_4$ singlets, as a result
of which there are no predictions on their mixing matrix, as it is
difficult to reconcile quarks with the neutrino sector~\footnote{The
  formulation of a full theory of flavour is beyond the scope of this
  work and will be left for the future.}.

The lepton and Higgs assignments are summarized in table\,\ref{tab1}.
\begin{table}[h!]
\begin{center}
\begin{tabular}{|c|c|c|c|c|c|c|c|c||c|c|}
\hline
&$\,L_e\,$&$\,L_{\mu}\,$&$\,L_{\tau}\,$&$\,\,l_{e}^c\,\,$&$\,\,l_{{\mu}}^c\,\,$&$\,\,l_{{\tau}}^c\,\,$&$N_{T}\,$&$\,N_4\,$&$\,\hat{H}\,$&$\,\eta\,$\\
\hline
$SU(2)$&2&2&2&1&1&1&1&1&2&2\\
\hline
$A_4$ &$1$ &$1^\prime$&$1^{\prime \prime}$&$1$&$1^{\prime \prime}$&$1^\prime$&$3$ &$1$ &$1$&$3$\\
\hline
\end{tabular}\caption{Summary of the relevant quantum numbers}\label{tab1}
\end{center}
\end{table}

The resulting leptonic Yukawa Lagrangian is:
\begin{eqnarray}\label{lag}
\mathcal{L}&=&y_e L_el_{_e}^c \hat{H}+y_\mu L_\mu l_{_\mu}^c \hat{H}+y_\tau L_\tau l_{_\tau}^c \hat{H}+\nonumber\\
&&+y_1^\nu L_e(N_T\eta)_{1}+y_2^\nu L_\mu(N_T\eta)_{1''}+y_3^\nu L_\tau(N_T\eta)_{1'}+\nonumber\\
&&+y_4^\nu L_e N_4 \hat{H}+ M_1 N_TN_T+M_2 N_4N_4+\mbox{h.c.}
\end{eqnarray}
This way the field $\hat{H}$ is responsible for quark and charged
lepton masses, the latter automatically diagonal.  The scalar
potential is:
\begin{eqnarray}\label{potential}
V&=&\mu_\eta^2\eta^\dagger\eta+\mu_{\hat{H}}^2 \hat{H}^\dagger \hat{H} 
+\lambda_1 [\hat{H}^\dagger \hat{H}]_1^2+\lambda_2 [\eta^\dagger\eta]_1^2
+\lambda_3 [\eta^\dagger\eta]_{1^\prime}[\eta^\dagger\eta]_{1^{{\prime}{\prime}}}\nonumber \\
&+&\lambda_4 [\eta^\dagger\eta^\dagger]_{1^\prime}[\eta\eta]_{1^{\prime\prime}}
+\lambda_{4^\prime}[\eta^\dagger\eta^\dagger]_{1^{\prime\prime}}[\eta\eta]_{1^\prime}
+\lambda_5[\eta^\dagger\eta^\dagger]_{1}[\eta\eta]_{1}
+\lambda_6([\eta^\dagger \eta]_{3_{1}}[\eta^\dagger \eta]_{3_{1}}+h.c.)\nonumber \\
&+&\lambda_7 [\eta^\dagger \eta]_{3_{1}}[\eta^\dagger \eta]_{3_{2}} 
+\lambda_8 [\eta^\dagger \eta^\dagger]_{3_{1}}[\eta \eta]_{3_{2}} 
+\lambda_9 [\eta^\dagger \eta]_{1^\prime}[\hat{H}^\dagger \hat{H}] 
+\lambda_{10}[\eta^\dagger \hat{H}]_{3_1}[\hat{H}^\dagger \eta]_{3_1} \nonumber\\
&+&\lambda_{11}([\eta^\dagger\eta^\dagger]_{1}\hat{H} \hat{H}+h.c.) 
+\lambda_{12}([\eta^\dagger\eta^\dagger]_{3_{1}}[\eta \hat{H}]_{3_1}+h.c.) 
+\lambda_{13}([\eta^\dagger\eta^\dagger]_{3_{2}}[\eta \hat{H}]]_{3_1}+h.c.) 
+\lambda_{14}([\eta^\dagger \eta]_{3_{1}}\eta^\dagger \hat{H}+h.c.) \nonumber\\
&+&\lambda_{15}([\eta^\dagger \eta]_{3_{2}}\eta^\dagger \hat{H}+h.c.) 
\end{eqnarray}
where $[...]_{3_{1,2}}$ is the product of two triplets contracted into
one of the two triplet representations of $A_4$, see
eq.\,(\ref{pr}),and $[...]_{1,1^\prime,1^\prime\prime}$ is the product
of two triplets contracted into a singlet representation of $A_4$.
In what follows we assume, for simplicity, CP conservation, so that
all the couplings in the potential are real.
For convenience we also assume $\lambda_4 = \lambda_4^\prime$ in order
to have manifest conservation of CP in our chosen
$A_4$ basis~\footnote{To see this, consider for instance the coupling
  $(\omega \lambda_4 +\omega^2 \lambda_4'
  )(\eta^\dagger_1\eta_2)^2+h.c.$ arising from the terms proportional
  to $\lambda_4$ and $\lambda_4'$ in eq.\,(\ref{potential}).  Since
  $\omega+\omega^2=-1$ this coupling is real if and only if $\lambda_4
  = \lambda_4^\prime$.}.

The minimization of the scalar potential results in :

\begin{equation}
  \vev{ H^0}=v_H\ne 0,~~~~ \vev{ \eta^0_1}=v_\eta \ne 0,~~~~
\vev{ \eta^0_{2,3}}=0\,,\label{vevs}
\end{equation}
where all vevs are real.  This vev alignment breaks the group $A_4$ to
its subgroup $Z_2$ responsible for the stability of the DM as well as
for the neutrino phenomenology~\cite{Hirsch:2010ru}.
In Appendix \ref{su5} we comment on a possible embedding of the model
into the grand unified group $SU(5)$.

\begin{center}{\it The stability of the DM}\end{center}

Since there are no couplings with charged fermions nor quarks because
of the $A_4$ symmetry, the only Yukawa interactions of the lightest neutral component
of $\eta_{2,3}$ are with the heavy \321 singlet right-handed neutrinos.
This state is charged under the $Z_2$ parity that survives
after the spontaneous breaking of the flavor symmetry.
One finds that, as a consequence of this $Z_2$ symmetry, the mass
matrix for the neutral scalars is block-diagonal, see
eq.\,(\ref{blockdiag}), so that the lightest neutral component of
$\eta_{2,3}$ is not mixed with the two Higgs scalars that take vev,
$H$ and $\eta_1$. Thus quartic couplings will not induce decays for
this DM candidate which is therefore stable and constitutes our DM
candidate.

We now show the origin of such a parity symmetry.  \vskip5.mm

As explained in Appendix \ref{a4group}, the group $A_4$, has two
generators: $S$, and $T$, that satisfy the relations
$S^2=T^3=(ST)^3=I$. In the three dimensional basis $S$ is given by 
\begin{equation}S=\left(
\begin{array}{ccc}
1&0&0\\
0&-1&0\\
0&0&-1\\
\end{array}
\right).\label{z2sim}\end{equation}
$S$ is the generator of the $Z_2$ subgroup of
$A_4$. The alignment $\vev{ \eta} \sim (1,0,0)$ breaks spontaneously $A_4$ to $Z_2$ since $(1,0,0)$ is 
manifestly invariant under the $S$ generator:
\begin{equation}S\vev{ \eta} =
  \vev{ \eta}\label{inv}.
\end{equation} For a generic triplet
irreducible representation of $A_4$, $\Psi=(a_1,~a_2,~a_3)^T$, we
have:
\begin{equation}S \Psi =\left(
\begin{array}{ccc}
1&0&0\\
0&-1&0\\
0&0&-1\\
\end{array}
\right)\left(\begin{array}{c}a_1 \\a_2\\a_3\end{array}\right)=\left(\begin{array}{c}a_1 \\-a_2\\-a_3\end{array}\right).\end{equation}
The $Z_2$ residual symmetry is defined as
\begin{equation}\label{residualZ2}
\begin{array}{lcrlcr}
N_1 &\to& +N_1\,,\quad& \eta_1 &\to& +\eta_1 ,\\  
N_2 &\to& -N_2\,,\quad& \eta_2 &\to& -\eta_2 ,\\
N_3 &\to& -N_3\,,\quad& \eta_3 &\to& -\eta_3 ,
\end{array}
\end{equation}
and the rest of the matter fields are $Z_2$ even, because the singlet
representation transforms trivially under $S$, see Appendix
~\ref{a4group}.
This is the residual symmetry which is responsible for the stability
of our DM candidate.

\begin{center}
{\it Neutrino phenomenology}  
\end{center}

  Here we summarize the main results obtained in
  Ref.\cite{Hirsch:2010ru} concerning the neutrino phenomenology.
The model contains four heavy right-handed neutrinos so it is a special
case, called (3,4), of the general type-I seesaw
mechanism~\cite{schechter:1980gr}.  After electroweak symmetry
breaking, it is characterized by Dirac and Majorana mass-matrix:
\begin{equation}
m_D=\left(
\begin{array}{cccc}
x_1&0&0&x_4\\
x_2&0&0&0\\
x_3&0&0&0\\
\end{array}
\right),\quad
 M_R=\left(
 \begin{array}{cccc}
 M_1&0&0&0\\
 0&M_1&0&0\\
 0&0&M_1&0\\
 0&0&0&M_2
 \end{array}
 \right).
\end{equation}
where $x_{1},x_2,x_3$ and $x_4$ are respectively proportional to
$y_1^\nu$, $y_2^\nu$, $y_3^\nu$ and $y_4^\nu$ of eq.\,(\ref{lag}) and
are of the order of the electroweak scale, while $M_{1,2}$ are assumed
close to the unification scale.
Light neutrinos get Majorana masses by means of the type-I seesaw relation and the light-neutrinos mass matrix
has the form:
\begin{equation}
\label{mnu}
m_\nu=-m_{D_{3\times 4}}M_{R_{4\times 4}}^{-1}m_{D_{3\times 4}}^T\equiv
\left(
\begin{array}{ccc}
y^2& ab&ac\\
ab&b^2&bc\\
ac&bc&c^2
\end{array}
\right).
\end{equation}
%
This texture of the light neutrino mass matrix has a null eigenvalue
$m_3=0$ corresponding to the eigenvector $(0,\,-b/c,\,1)^T,$
\footnote{ Note that if we were to stick to the minimal
  (3,3)-type-I seesaw scheme, with just 3 SU(2) singlet states, one
  would find a projective nature of the effective tree-level light
  neutrino mass matrix with two zero eigenvalues, hence
  phenomenologically inconsistent. That is why we adopted the (3,4)
  scheme.}  implying a vanishing reactor mixing angle $ \theta_{13}=0$
and inverse hierarchy.  The atmospheric angle, the solar angle and the
two square mass differences can be fitted.  
The model implies a neutrinoless double beta decay effective mass parameter in the range 0.03 to 0.05~eV at
3~$\sigma$, within reach of upcoming experiments.

\begin{center}{\it Notation}\end{center}

After electroweak symmetry breaking and the minimization of the potential we can write:
\begin{equation}\begin{array}{cc}
 \hat{H}=\left(
\begin{array}{c}
H^{\prime +}_0\\
(v_H+H^{\prime}_0+i A^{\prime}_0)/\sqrt{2}
\end{array}
\right),
&\eta_1=\left(
\begin{array}{c}
H^{\prime +}_1\\
(v_\eta+H'_1+iA'_1)/\sqrt{2}
\end{array}
\right),\\ \\
\eta_2=\left(
\begin{array}{c}
H^{\prime +}_2\\
(H'_2+iA'_2)/\sqrt{2}
\end{array}
\right),&\eta_3=\left(
\begin{array}{c}
H^{\prime +}_3\\
(H'_3+iA'_3)/\sqrt{2}
\end{array}
\right).
\end{array}
\end{equation}

The structure of the neutral and charged scalar mass matrices follows
from the exact $Z_2$ symmetry, which forbids mixings between particles
with different $Z_2$ parities and from CP conservation. 
As a result the charged and neutral scalar mass matrices decompose
into two-by-two mass matrices which, after diagonalization, give the
scalar mass spectrum of the model.
In what follows, unprimed particles denote mass eigenstates. We refer
to Appendix \ref{spectrum} for details about the mass matrices and for
a complete description of the mass spectrum.
The $Z_2$-odd sector contains two real CP even scalars, $H_2$ and
$H_3$, two real CP odd scalars, $A_2$ and $A_3,$ and four charged
scalars, $H_2^{\pm}$ and $H_3^{\pm}$.  The $Z_2$ even scalars consist
of two real CP even scalars $H$ and $H_0$, that we generically call
'Higgses', a pseudoscalar $A_0$ and two charged scalars $H^{\pm}_0.$
The masses of the $W^{\pm}$ and $Z$ gauge bosons impose the relation
$v_H^2+v_\eta^2=v^2$ where $v$ stands for the standard model value of 
the vev.  We call $\tan\beta$ the ratio of the two vevs:
$\tan\beta=v_H/v_\eta$.

The DM candidate of the model corresponds to the lightest $Z_2$-odd
neutral spin zero particle which, for the sake of definiteness, we
take as the CP-even state $H_2$.  We remind that the
  parameters of the model relevant for the DM phenomenology are the 15
  couplings of the scalar potential, $\lambda_{i},$ and the ratio of
  the vacuum expectation values $\tan \beta= v_{H}/v_{\eta}.$ Indeed
  the minimization of the scalar potential and electroweak symmetry
  breaking allow to recast the mass parameters $\mu_H$ and
  $\mu_{\eta}$ in terms of the couplings $\lambda_{i}$ and $\tan\beta.$
Note that the couplings and the Majorana
  mass parameters in Eq. Ref.\ref{lag} determine the neutrino mass
  matrix, they are not relevant for the dark matter phenomenogy. 
Before moving to the calculation of its relic abundance in the next
section we consider the phenomenological constraints on the parameter
space of the model.

\section{Phenomenological constraints}
\label{constraints}

In order to find the viable regions in parameter space where to
perform our study of dark matter, we must impose the following
constraints to the model :

\begin{itemize}
\item {\it Electroweak precision tests} 

  \noindent 
  It is well-known that the oblique parameters $S,T,U$ provide
  stringent constraints on theories beyond the Standard
  Model~\cite{Peskin:1991sw}.
  Concerning the $S$ and $U$ parameters, these receive negligible
  contributions from the scalars of the model
  \cite{Grimus:2008nb,Barbieri:2006dq}, hence we focus on the $T$
  parameter.  We compute the effect on $T$ induced by the scalars
  following \cite{Grimus:2007if} and we impose the bounds from
  electroweak measurements \cite{Nakamura:2010zzi}:
$$-0.08\leq T \leq0.14.$$
While this bound favors a light Standard Model Higgs boson, large
Higgs masses are possible in the presence of new physics, such as our
multi-Higgs-doublet model.
Indeed, a negative deviation of the electroweak $T$ parameter induced
by a heavy Higgs can be compensated by a positive $\Delta T$ produced
by new scalar particles of the model, so one can raise the Higgs mass
up to $\sim600$ GeV or so (see below)~\footnote{A similar situation
  holds, for instance, in the Inert Doublet
  Model~\cite{Barbieri:2006dq}.}.

  We have explicitly verified that we can choose the mass spectrum of
  the model in such a way that this constraint is always respected. We
  refer to Appendix \ref{oblique} for more detail.

\item {\it Collider bounds}

  \noindent Searches for supersymmetric particles at LEP place a lower
  bound on the chargino mass of $\sim$ 100 GeV. 
  To be conservative we apply this constraint also to the masses of
  the charged scalars of our model, even if slightly lower masses,
  $70-90$ GeV, might still be consistent with LEP
  data\cite{Pierce:2007ut}.

  The bounds imposed by LEP II on the masses of the neutral scalars in
  our model are similar to those constraining the Inert Doublet Model,
  given in Ref.\cite{Lundstrom:2008ai}.  This analysis applies
  directly to the $Z_2$-odd scalar sector of our model of particular
  interest for DM phenomenology, as it constrains the mass difference
  between $H_{2}$ and $A_{2}$.
  The excluded region that we adopt is taken from Fig.7 of
  Ref.\cite{Lundstrom:2008ai}.

  The masses of the $Z_2$ even neutral scalars are also constrained by
  LEP searches.  Although a precise bound requires a detailed analysis
  of the even sector, we just impose a lower limit on these masses of
  114 GeV, which is approximatively the LEP limit on the SM higgs
  mass.

  We remark that lepton flavor violating processes are suppressed by
  the large right-handed neutrino scale.

\item {\it Perturbativity and vacuum stability}

  \noindent The requirement of perturbativity  imposes the following bounds on the Yukawa couplings of the model
  and $\tan\beta:$
\begin{eqnarray}
\lambda_i \lsim \sqrt{4 \pi} \hspace{0.5cm}\mbox{i=1,...,15}\hspace{0.05cm},\nonumber \\
\tan\beta >0.5\hspace{0.5cm} \nonumber
\end{eqnarray}
this leads to an upper bound on the masses of the scalars at $\sim600$
GeV~\footnote{Note that this bound would not apply to the Inert Higgs Doublet model, which would
    potentially allow heavier dark matter masses.}
The constraint on $\tan\beta$ is necessary in order to
preserve a perturbative top-Yukawa coupling.

Finally we must impose the stability of the vacuum. The conditions
ensuring that the potential is bounded from below are:
$$\lambda_1>0 \hspace{0.05cm},  \hspace{0.5cm} \lambda_2+\lambda_3+2\lambda_4+\lambda_5>0 \hspace{0.05cm},$$
$$\lambda_1+3(\lambda_2+\lambda_3+2\lambda_4+\lambda_5)+3(\lambda_9+Min1)+
3(2\lambda_2-\lambda_3+\lambda_8+Min2)-6Q >0,$$

where $$Min1=Min(\lambda_{10}-2|\lambda_{11}|,0),$$ $$Min2=Min(\lambda_7-2|\lambda_4-\lambda_5-\lambda_6|,0),$$ 
$$Q=|\lambda_{12}|+|\lambda_{13}|+|\lambda_{14}|+|\lambda_{15}|.$$

\end{itemize}
\begin{figure*}[t]
\includegraphics[width=0.32\textwidth]{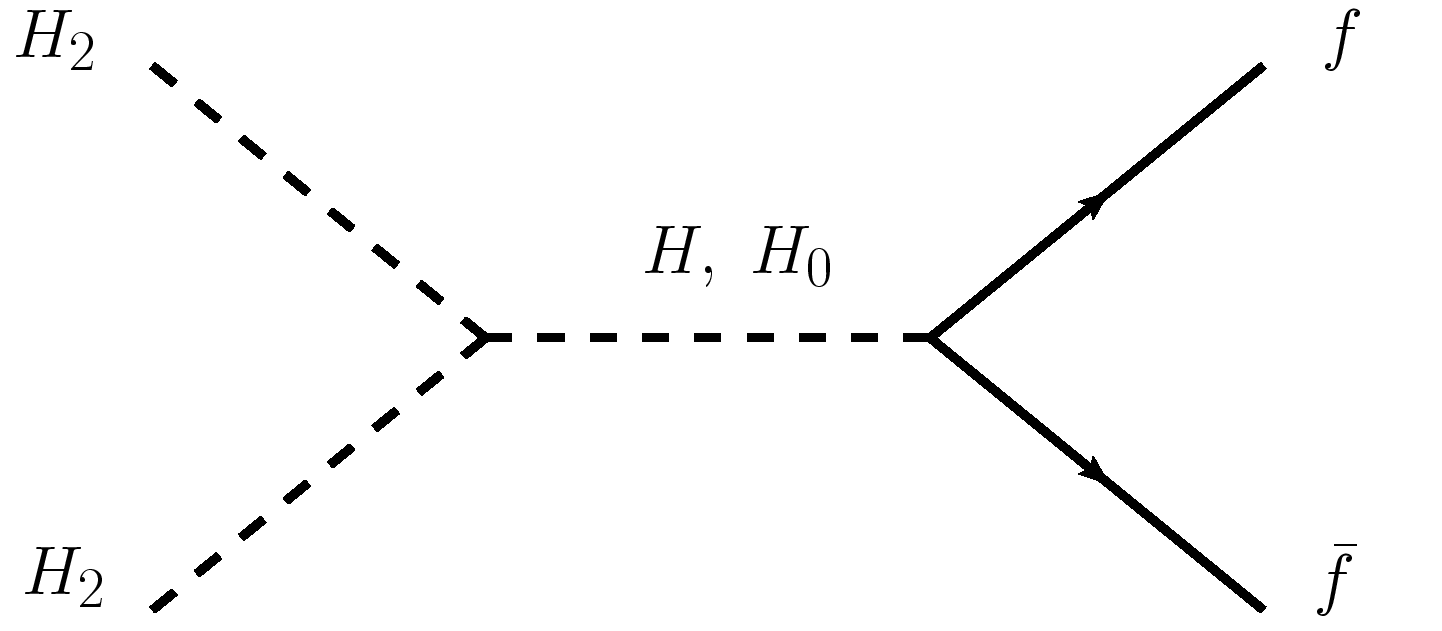}
\includegraphics[width=0.32\textwidth]{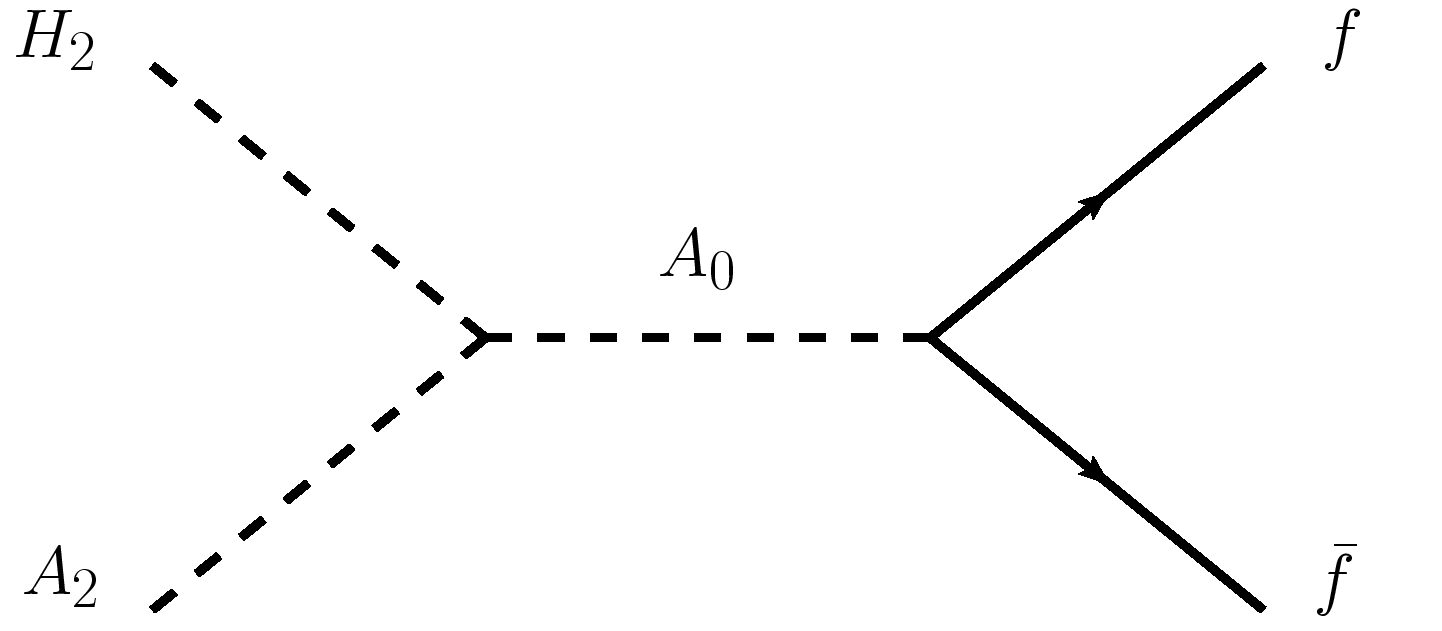}
 \includegraphics[width=0.32\textwidth]{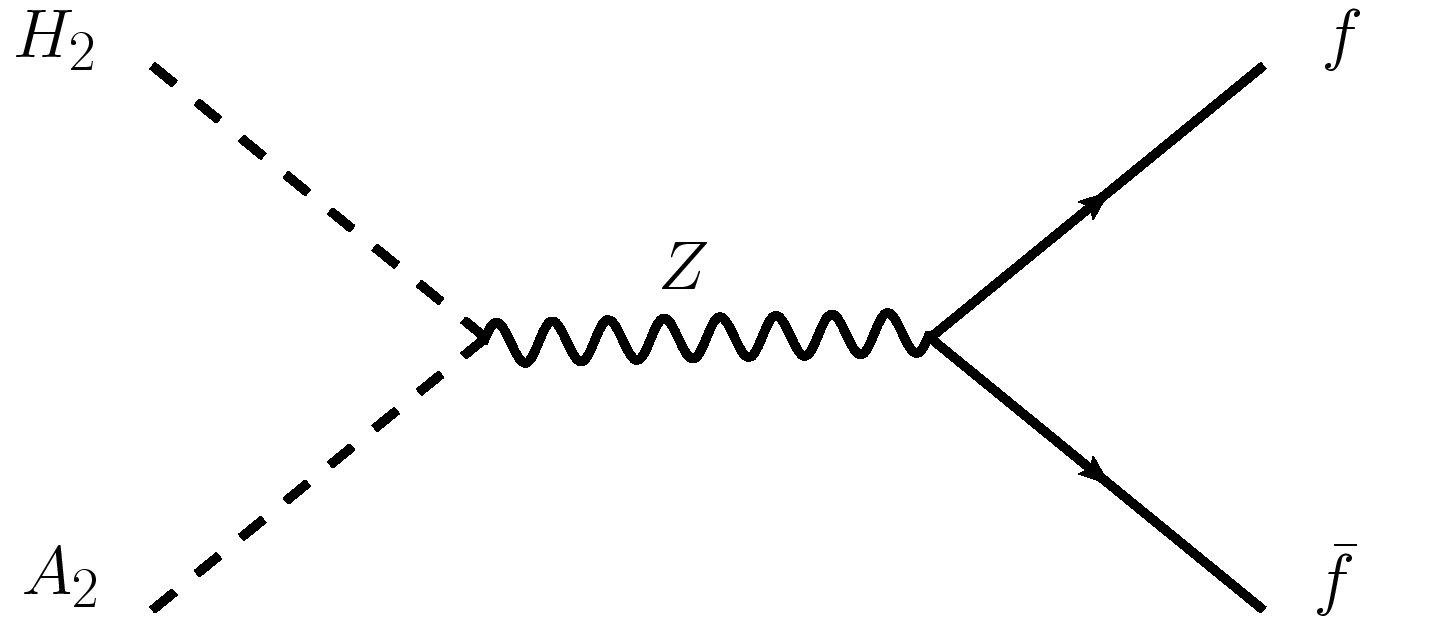}
 \caption{Left: Tree level Feynman diagrams for the $H_2$
   annihilation into fermions. Center and right: diagrams for $H_2$
   co-annihilation with the pseudscalar $A_2$ into fermions.}
  \label{diagrams}
\end{figure*}
\begin{figure*}[t]
\includegraphics[width=0.33\textwidth]{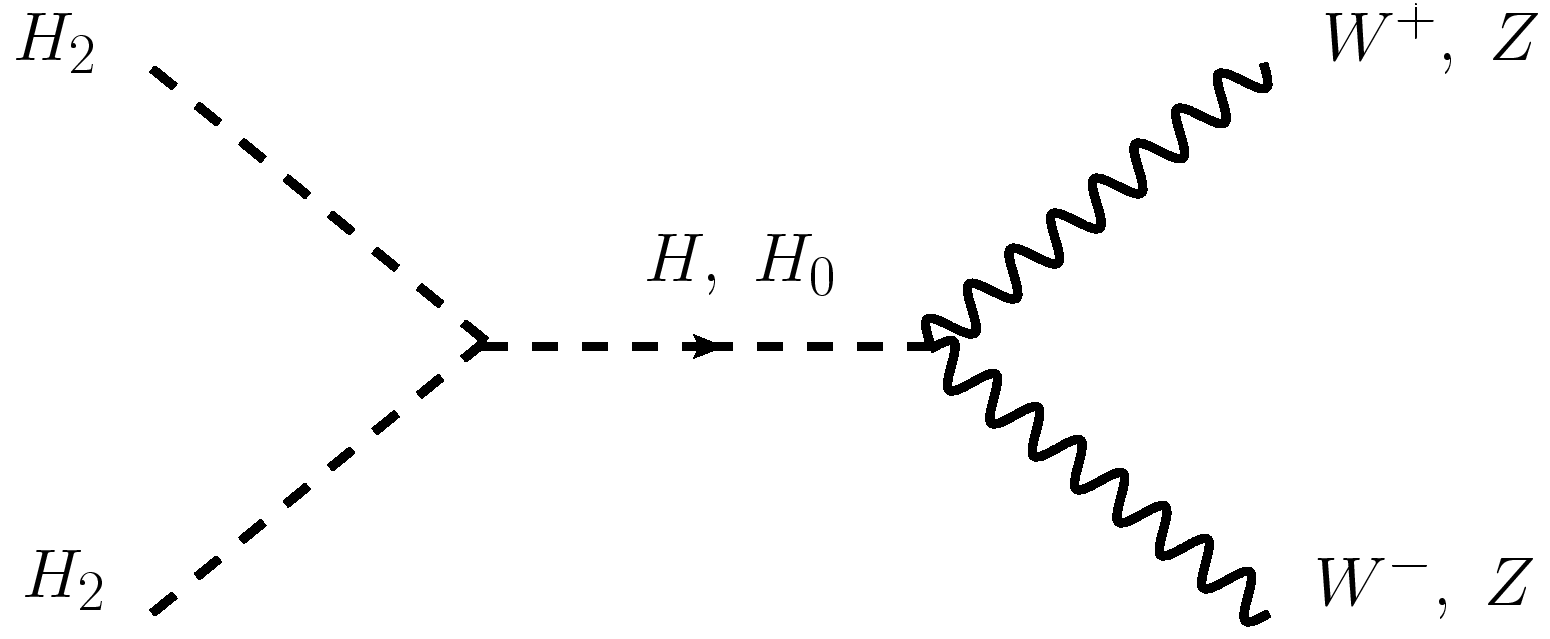}
\includegraphics[width=0.20\textwidth]{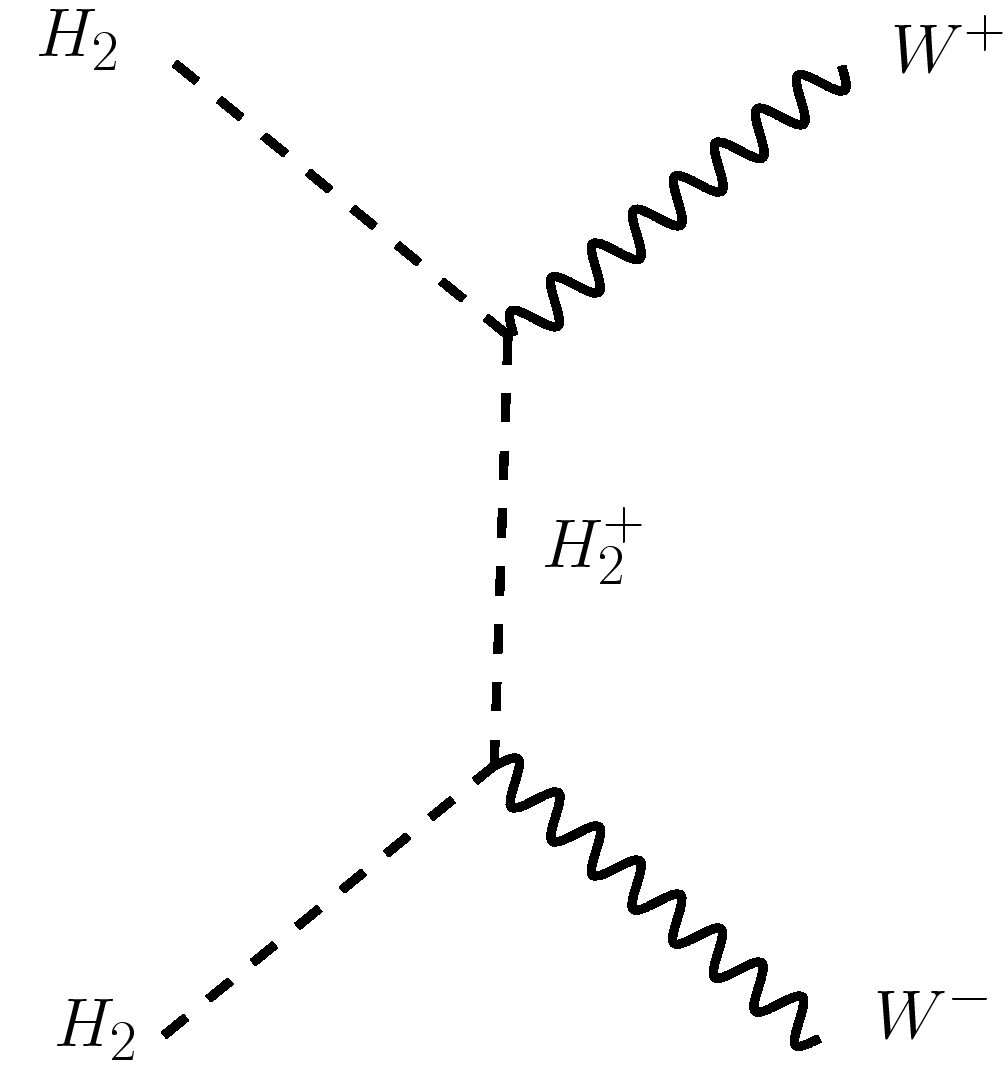}
 \includegraphics[width=0.25\textwidth]{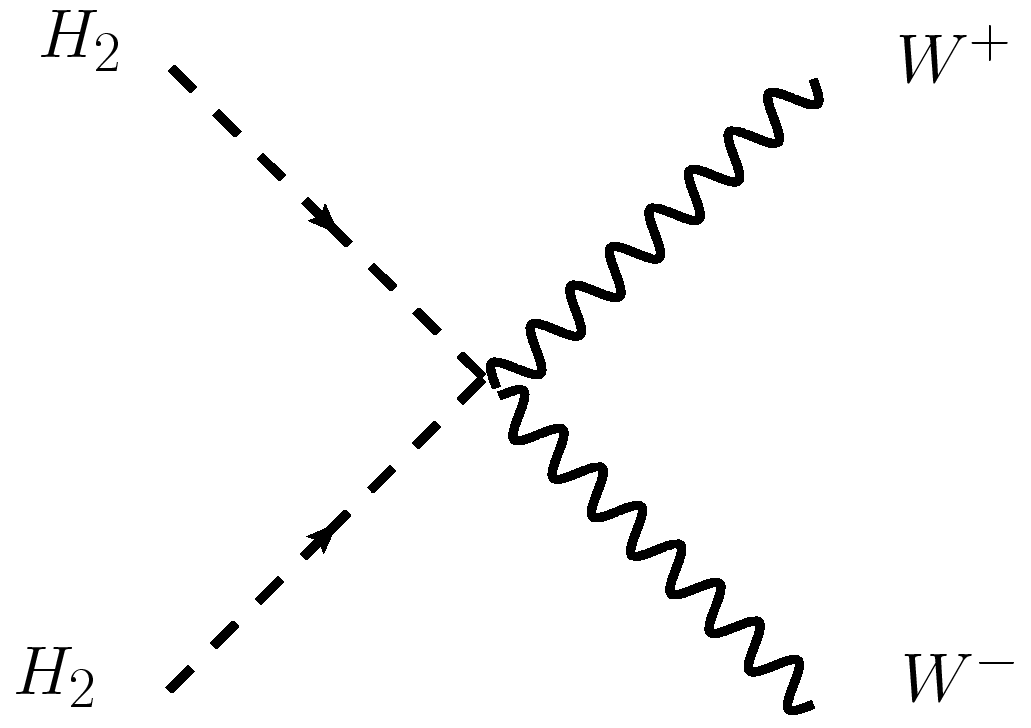}
 \includegraphics[width=0.20\textwidth]{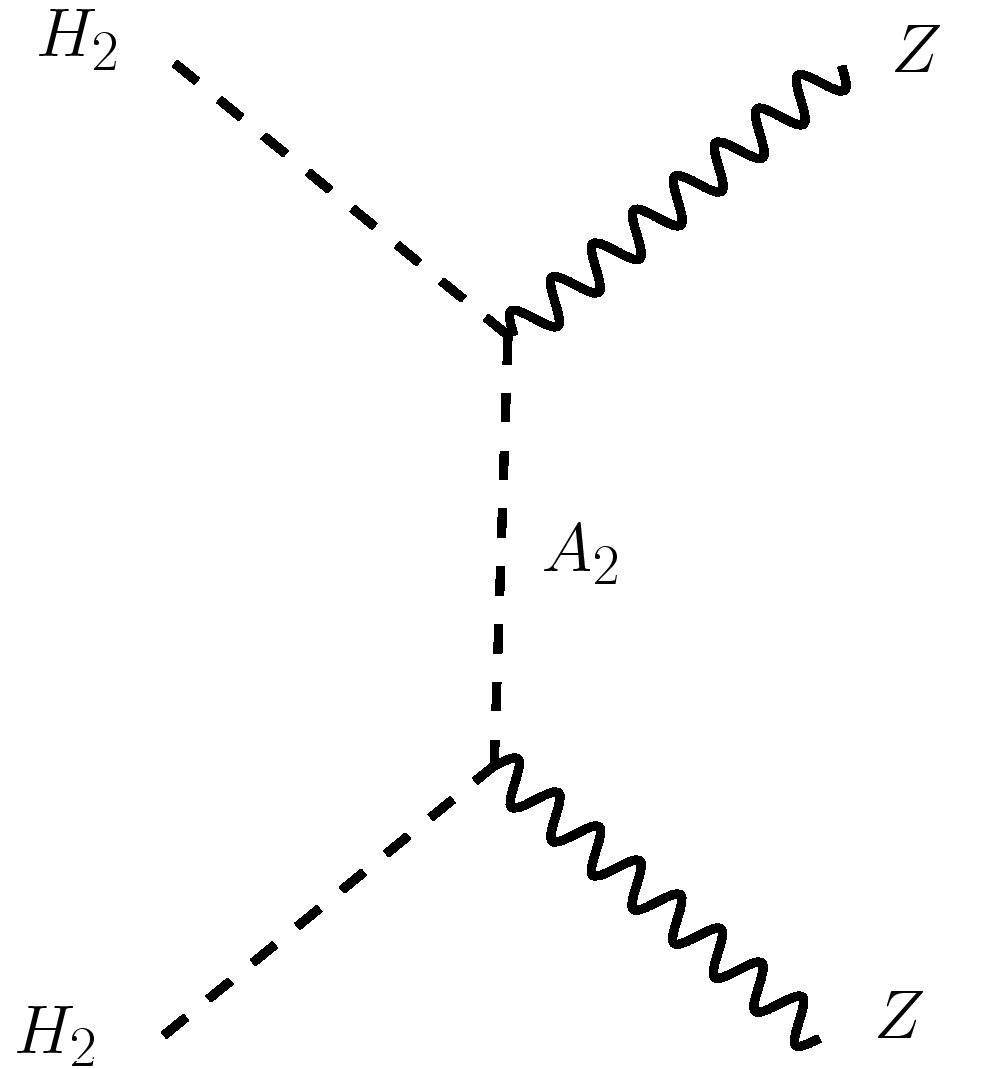}
  \caption{Tree level Feynman diagrams for the $H_2$ annihilation into $W^{\pm}$ and $Z$ $Z$.}
  \label{diagramsw}
\end{figure*}

\section{Relic Density}
\label{relic}

\begin{center}
\begin{figure}[t]
\includegraphics[width=0.4\textwidth]{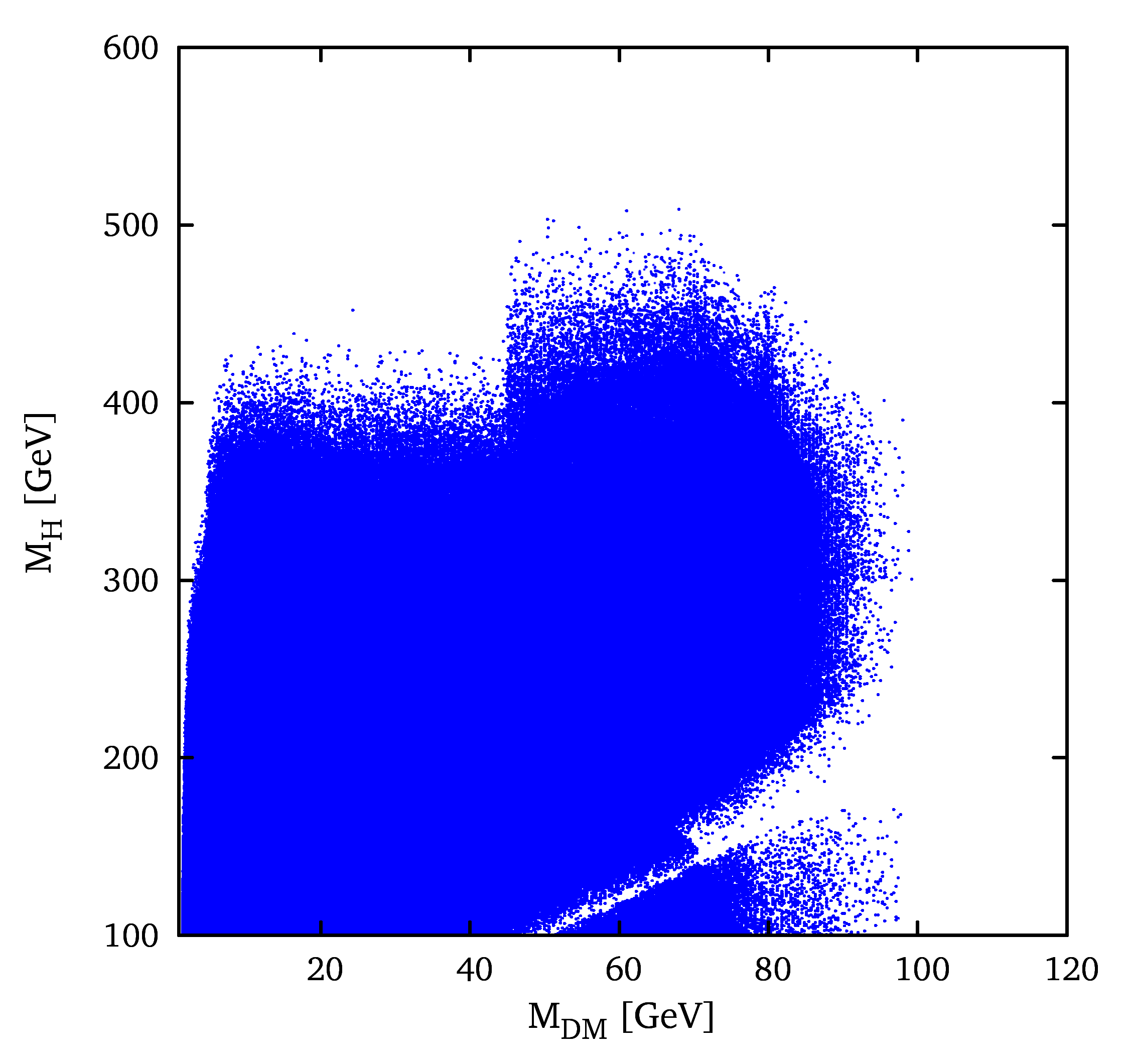}
\caption{Regions in the plane DM mass ($M_{DM}$) - lightest Higgs boson $M_H$
    allowed by collider constraints and leading to a DM relic abundance
  compatible with WMAP measurements.}
  \label{omega}
\end{figure}
\end{center}
The thermal relic abundance of $H_2$ is controlled by its annihilation
cross section into SM particles.  In Fig.\ref{diagrams} and
Fig.\ref{diagramsw} we show the Feynman diagrams for the most relevant
processes.  In order to study the viable regions of the model we
perform a random scan over the 16-dimensional parameter space
($\lambda_i$ and $\tan \beta$ ) and compute the dark matter relic
abundance using the micrOMEGAs
package\cite{Belanger:2010gh,Belanger:2008sj}.  
  Using the mass relations in equations \ref{mass_eq} we can trade
  some of the couplings $\lambda_i$ for the scalar masses. This system
  of 10 linear equations allows us to trade only 8 of the $\lambda_i
  .$ Indeed, the texture of the mass matrix of the $Z_2$ odd sector
  imposes the following constraints on the squared mass differences:
  $M_{H_2}^2-M_{H_3}^2=3
  (M_{A_2}^2-M_{A_3}^2)=3(M_{H_2^+}^2-M_{H_3^+}^2).$ We linearly
  sample on the 8 independent masses in such a way to fulfill the
  collider constraints discussed in Sec.\ref{constraints}. The
  remaining 7 free $\lambda_{i}$ and $\cos \beta$ are linearly sampled
  inside their allowed ranges, specified in Sec.\ref{constraints}.
  Then, we select those choices which satisfy the perturbativity
  constraints for all the 15 $\lambda_i,$ the vacuum stability bounds
  and for which the electroweak precision tests are satisfied.
Finally, we choose only those models which provide a relic adundance
consistent with the WMAP measurements:
$$0.09\leq\Omega h^2 \leq 0.13.$$

In Fig.~\ref{omega} we show the regions with a correct relic abundance
in the plane DM mass ($M_{DM}$) and the lightest Higgs boson mass
$M_{H}$.  For dark matter masses well below the $W^{\pm}$ threshold,
dark matter annihilations into fermions are driven by the s-channel
exchange of the Higgs scalars of the model, as shown in
Fig.~\ref{diagrams}.

For $M_{H}\lesssim 400$ GeV the annihilation cross-sections are large
enough to obtain the correct relic density for all DM masses up to the
the $W^{\pm}$ threshold. At larger Higgs boson masses, annihilations
into light fermions are suppressed so that the relic abundance is
typically too large unless efficient co-annihilations with the
pseudoscalar $A_2$ or with $H_2^{\pm}$ occur.  These processes, shown
in Fig.\ref{diagrams} for $A_2$, take place only for small mass
splittings between $H_2$ and $A_2$ or $H_2^{\pm}$.  Note that the
possibility to co-annihilate with the charged scalar is ruled out,
since LEP data requires $M_{H_2}^{\pm} \gsim 100$ GeV. There is,
however, a narrow window still allowed by LEPII limits on the
$A_2$-$H_2$ plane\cite{Lundstrom:2008ai} where arbitrarily small mass
splitting can exist between $H_2$ and $A_2$ for $M_{DM}\gsim 40$ GeV.
For lighter dark matter particles, strong co-annihilations can not be
reconciled with LEPII constraints as these require $M_{A_2}\gsim 100$
GeV \cite{Lundstrom:2008ai}.
We can therefore exclude by cosmological observations the parameter
region correspondig to (simultaneously) $M_{DM}\lesssim 40$ GeV and 
$M_{H}\gsim 400$
GeV, because the DM would be overabundant.  This is clearly seen in
Fig.~\ref{omega}.
In contrast, the absence of points on the strip corresponding to the
line $M_{DM}\sim M_H/2$ is associated to the presence of the $H$
resonance, which would enhance the DM annihilation cross section
giving a too small Dark Matter abundance.

For dark matter masses larger than $M_W$, unless the dark matter
candidate is very heavy $M_{DM}\gsim 500$ GeV, the annihilation cross
section into gauge bosons is typically too large so that $H_2$ can
only be a subdominant component of the dark matter budget of the
Universe. 
However, for certain combinations of masses and parameters, the
annihilations into gauge bosons may be suppressed by a cancellation
between the Feynman diagrams (Fig.\ref{diagramsw}), leading to an
acceptable relic density. Indeed, this happens for some points in
Fig.~\ref{omega}~\footnote{For the Inert Doublet model, these
  cancellations have been noted in \cite{Lundstrom:2008ai} and have
  been recentely studied in detail in
  \cite{LopezHonorez:2010tb}.}. Ref.~\cite{LopezHonorez:2010tb} shows
that these processes allow for dark matter masses up to $160$ GeV,
just below the threshold for annihilations into top pairs.  In our
scan, the viable region of the parameter space extends only up to
$M_{DM}\sim 100$ GeV. Solutions at higher masses cannot be excluded,
though their scrutiny is rather involved due to the large number of
parameters of the model.


We now turn to the region $M_{DM}\sim 500$ GeV.  As noted in
\cite{Hirsch:2010ru}, a scalar dark matter candidate annihilating into
massive vector bosons inherits the correct relic abundance for this
value of the mass if all other annihilation processes are absent.
This scenario could be realised in our model by tuning the Yukawa
couplings in order to suppress the dark matter annihilation into Higgs
scalars and fermions.  However, since the dark matter mass arises
entirely from the electroweak symmetry breaking sector, high dark
matter masses tend also to require large scalar couplings.  This
argument suggests that solutions with a heavy dark matter candidate
and small couplings with the other scalars would be rather fine-tuned
and we do not consider this possibility any further\footnote{ Note
  also that in the range $\sim60$ GeV-$M_W,$ annihilations into three
  body final states may play a role, as shown for the Inert Doublet
  model in \cite{Honorez:2010re}.}.
In the next two sections we study the prospects for direct and
indirect dark matter detection.

\section{Direct detection}
\label{direct}

\begin{center}
\begin{figure*}[t]
\includegraphics[width=0.4\textwidth]{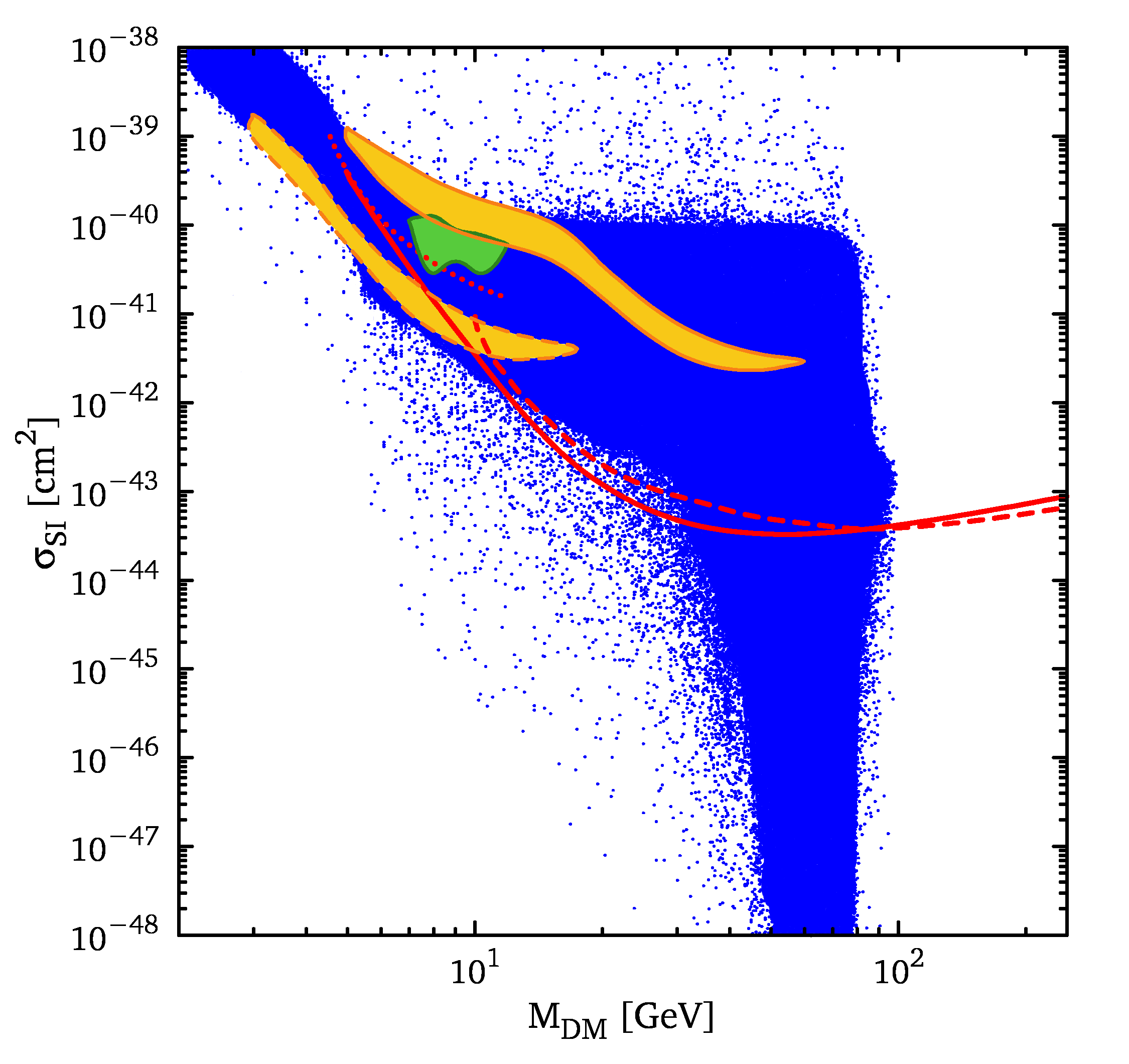}
\includegraphics[width=0.4\textwidth]{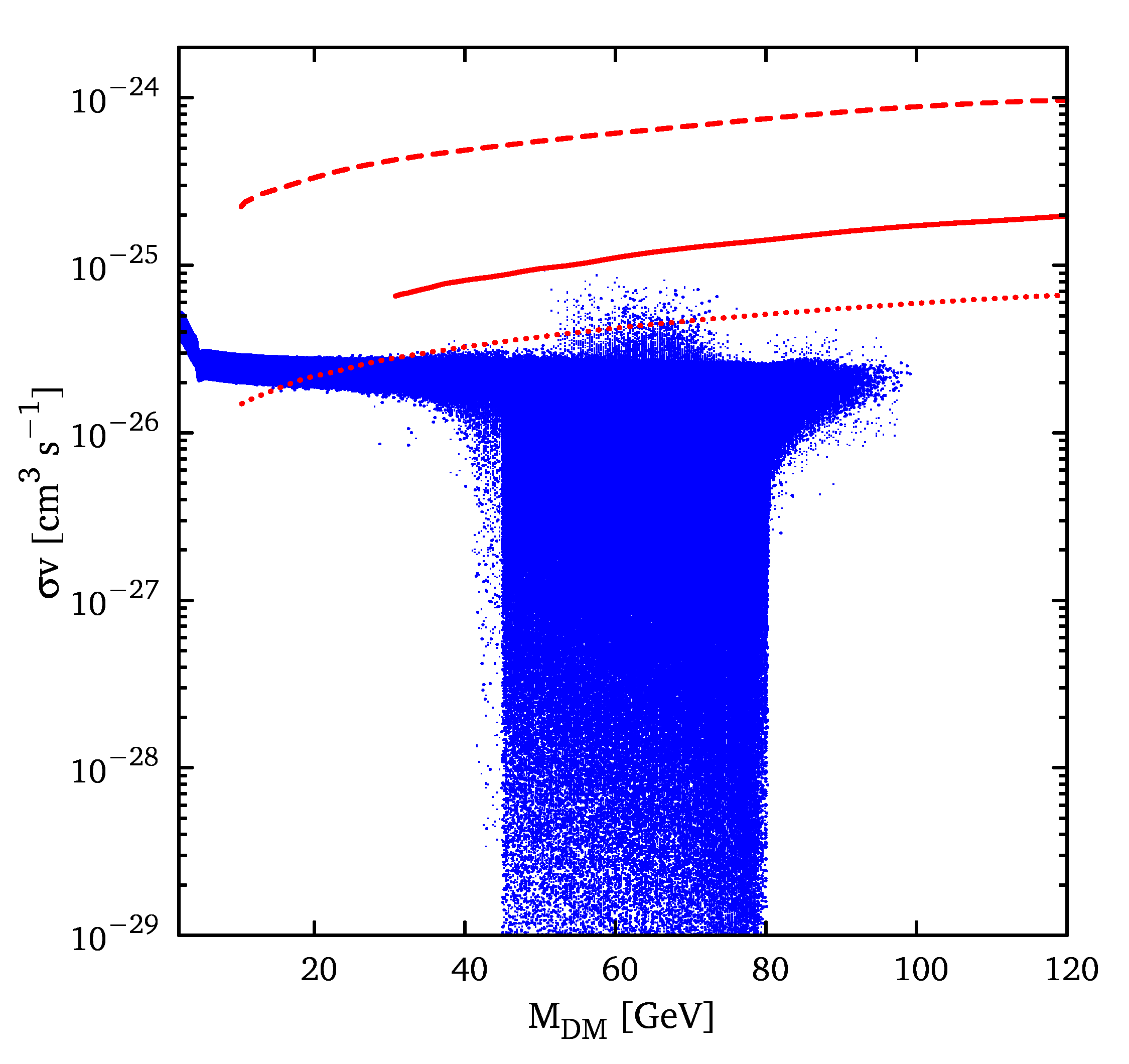}
  \caption{Left plot: Spin-independent DM scattering cross section 
off-protons as a function of the dark matter mass. The orange regions 
delimited by the dashed (solid) line show the DAMA/LIBRA annual modulation 
regions including (neglecting) the channeling effect~\cite{Fornengo:2010mk}.
The green region corresponds to the COGENT data~\cite{Savage:2010tg}.
Dashed and dotted red lines correspond to the upper bound from CDMS
(respectively from \cite{Ahmed:2009zw} and \cite{Ahmed:2010wy}).
XENON100 bounds \cite{Aprile:2010um} are shown as a solid red line.
Right plot: annihilation cross section times velocity as a function of
the dark matter mass. The solid and dashed red lines show respectively
the upper bound inferred by Fermi-LAT observations of the Draco dwarf
galaxy \cite{Abdo:2010ex} and FERMI-LAT measurements of the isotropic
diffuse gamma-ray emission \cite{Abdo:2010dk}.  Projected 5 years
sensitivity from measurements of the isotropic diffuse gamma-ray
emission are shown as a dotted red line \cite{Abazajian:2010zb}.}
  \label{dd-id}
\end{figure*}
\end{center}
Dark Matter can be searched for in underground detectors looking for
nuclear recoils induced by dark matter scattering against the target
material.  The scalar dark matter candidate we are considering couples
to quarks via Higgs boson exchange, leading therefore to pure spin
independent (SI) interactions with the nucleons.  In the left panel of
Fig.\ref{dd-id} we show the SI scattering cross section off proton for
the models with a correct dark matter abundance.  We note that a large
region of the parameter space is ruled out by the constrained imposed
by current dark matter direct detection experiments.

A positive signal of dark matter detection has been claimed by the
DAMA collaboration \cite{Bernabei:2000qi}.  DAMA has reported a high
statistical evidence for annual modulation of the event rate over 13
year cycles \cite{Bernabei:2000qi,Bernabei:2010mq}.  These results
have prompted many attempts to interpret the data in terms of dark
matter interactions with nuclei. Assuming an elastic WIMP interactions
off nuclei and for "standard" astrophysical assumption on the local DM
density and velocity distribution, the DAMA signal is in conflict with
the null results reported by other
experiments\cite{Savage:2010tg,Kopp:2009qt,Andreas:2010dz}. However
astrophysical inputs are subject to large uncertainties (see
e.g.\cite{Belli:1999nz,Green:2010gw,Fox:2010bz}) and, moreover, the
detector response is not completely known (for example the role of
channelling is still under debate \cite{Bozorgnia:2010zc}).  We also
note that there exist alternative particle physics scenarios where a
compability between the DAMA signal and the results of other
experiments can be obtained,
e.g.\cite{Barger:2010gv,TuckerSmith:2001hy,Chang:2009yt}.

Recently, a hint of a possible signal from dark matter has been reported by
the COGENT experiment \cite{Aalseth:2010vx}.
In Fig.\ref{dd-id} we show the combination of SI scattering cross section and DM masses
wich can fit the DAMA and COGENT results.
An excess of events over the expected background has also been found by the CDMS experiment \cite{Ahmed:2009zw}, 
although with a low statistical
significance and by CRESST \cite{Cresst}, even if these results are
still preliminary.  Interestingly, all these possible signals point to
the same region of the parameter space, favoring low dark matter
masses $M_{DM}\lesssim 15$ GeV
\cite{Savage:2010tg,Kopp:2009qt,Schwetz:2010gv,Fornengo:2010mk}.  As seen in
Fig.\ref{dd-id}, these possible hints of dark matter detection can be
accomodated within our model.

Current upper bounds severely challenge a possible interpretation of
the aforementioned anomalies in terms of dark matter SI interactions.
Indeed, as shown in Fig.\ref{dd-id}, the bulk of the COGENT and DAMA
regions are excluded by the constraints inferred by XENON100
\cite{Aprile:2010um} and CDMS \cite{Ahmed:2009zw,Ahmed:2010wy}.
Still, experimental uncertainties may significantly affect the upper
limits obtained by direct detection experiments, expecially for low
mass WIMPs (see \cite{Savage:2010tg,Schwetz:2010gv,Collar:2010ht}).
For this reason a possible dark matter spin-independent interpretation
of the excess, even if disfavored, cannot be excluded at present.

\section{Indirect detection}
\label{indirect}

Dark Matter indirect detection experiments look for signatures of DM
annihilation into photons, neutrinos and (anti-)~matter fluxes.  The
expected DM signals depend on the astrophysical details related to the
DM density distribution in the region of observation. Particle physics
enters in the determination of the DM mass, annihilation cross section
$\sigma v$ and the rates into various annihilation channels.  In
Fig.~\ref{dd-id} we show $\vev{\sigma v}$ at small velocity, relevant
for DM annihilations inside our galaxy, as a function of $M_{DM}$. For
$M_{DM}\lesssim 40$ GeV $\vev{\sigma v}$ remains close to the thermal
value at freeze-out, $3\times 10^{-26} \mbox{ cm}^3\mbox{s}^{-1},$ as
expected for the s-wave DM annihilation into light fermions.  At
larger DM masses, the presence of co-annihilations allows for much
smaller values of $\vev{\sigma v}$.  The solutions at $\vev{\sigma v}$
$ \sim 10^{-25} \mbox{ cm}^3\mbox{s}^{-1}$ correspond to DM masses
just below the Higgs resonance. In this case the annihilation cross
section at small velocities is boosted with respects to its values at
the DM freeze-out.  This behaviour of $\vev{\sigma v}$ close to a
narrow Breit-Wigner resonance has been recently widely exploited in
order to boost the annihilation signal so to explain the cosmic-rays
anomalies reported by the PAMELA collaboration
\cite{Ibe:2008ye,Guo:2009aj,Feldman:2008xs,Ibe:2009dx}.

In order to sketch the prospects for indirect DM detection we show in
Fig.~\ref{dd-id} the constraints on $\langle\sigma v\rangle$ imposed
by the Fermi-LAT observations of the Draco dwarf spheroidal galaxy
\cite{Abdo:2010ex} and the Fermi-LAT measurements of the isotropic
diffuse gamma-ray emission \cite{Abdo:2010dk}.  We caution that these
upper bounds have been computed assuming DM annihilations into $b
\bar{b}$, therefore they would directly apply only for parameter
choices in our model where this annihilation channel dominantes.
Still, this happens in large part of the parameter space, and in
particular at low dark matter masses, where the Fermi-LAT constraints
are close to the predictions of the model.  For a comparison of these
bounds with those obtained for different annihiliation channels we
refer the reader to the original references.  Further constraints for
different targets of observations are obtained in
Ref.~\cite{Ackermann:2010rg,Papucci:2009gd,Cirelli:2009dv,Zaharijas:2010ca,Anderson:2010hh}

One sees from Fig.~\ref{dd-id} that current bounds are not yet able to
significantly constrain the model. However, the Fermi-LAT sensitivity
is expected to improve considerably with larger statistics and for
different targets of observations, see
e.g. \cite{Abazajian:2010zb,Cuoco:2010jb,Zaharijas:2010ca,Anderson:2010hh}).
For example, in Fig.~\ref{dd-id} we show the forecasted 5 years
FERMI-LAT sensitivities from the isotropic diffuse gamma-ray emission
\cite{Abazajian:2010zb}.  Fermi-LAT measurements should be able to
test the model for low dark matter masses.

\section{ Conclusions and discussion}
\label{conclusions}

We have studied a model where the stability of the dark matter
particle arises from a flavor symmetry.  The $A_4$ non-abelian
discrete group accounts both for the observed pattern of neutrino
mixing as well as for DM stability.
We have analysed the constraints that follow from electroweak
precision tests, collider searches and perturbativity.
Relic dark matter density constraints exclude the region of the
parameter space where simultaneously $M_{DM}\lesssim 40$ GeV and
$M_{H}\gsim 400$ GeV because of the resulting over-abundance of dark
matter.
We have also analysed the prospects for direct and indirect dark
matter detection and found that, although the former already excludes
a large region in parameter space, we cannot constrain the mass of the
DM candidate.
In contrast, indirect DM detection is not yet sensitive enough to
probe our predictions. However, forecasted sensitivities indicate that
Fermi-LAT should start probing them in the near future.

All of the above relies mainly on the properties of the scalar sector
responsible for the breaking of the gauge and flavour symmetry.
A basic idea of our approach is to link the origin of dark matter  to
the origin of neutrino mass and the understanding of the pattern of
neutrino mixing, two of the most oustanding challanges in particle
physics today. At this level one may ask what are the possible tests
of this idea in the neutrino sector. 
Within the simplest scheme described in Ref.~\cite{Hirsch:2010ru} one
finds an inverted neutrino mass hierarchy, hence a neutrinoless double
beta decay rate accessible to upcoming searches, while $\theta_{13}=0$
giving no CP violation in neutrino oscillations.
Note however that the connection of dark matter to neutrino properties
depends strongly on how the symmetry breaking sector couples to the
leptons.

\section{Acknowledgments}

This work was supported by the Spanish MICINN under grants
FPA2008-00319/FPA and MULTIDARK Consolider CSD2009-00064, by
Prometeo/2009/091, by the EU grant UNILHC PITN-GA-2009-237920.
S.~M. is supported by a Juan de la Cierva contract. E.~P. is supported
by CONACyT (Mexico).

\appendix

\section{The $A_4$ group}
\label{a4group}

All finite groups are completly characterized by means of a set of
elements called generators of the group and a set of relations, so
that all the elements of the group are given as product of the
generators. The group $A_4$ consists of the even permutations of four
objects and then contains $4!/2=12$ elements.  The generators are $S$
and $T$ with the relations $S^2=T^3=(ST)^3=\mathcal{I}$, then the
elements are $1,S,T, ST, TS, T^2,ST^2,STS,TST,T^2S,TST^2,T^2ST$.

\begin{table}[t]
\begin{center}
\begin{tabular}{|l|c|c|c|c|}
\hline 
&$C_1=\{I\}$ & $C_2=\{T\}$ & $C_3=\{T^2\}$ & $C_4=\{S\}$ \\
\hline
1&1 & 1&1 &1\\
\hline
$1'$&1 &$\omega$ & $\omega^2$&1\\
\hline
$1''$&1 &$\omega^2$ & $\omega$&1\\
\hline
3&3 &0 & 0&$-1$\\
\hline
\end{tabular}\caption{Character table of $A_4$ where $C_i$ are the different classes and $\omega^3\equiv 1$. }\label{tab0}
\end{center}
\end{table}

$A_4$ has four irreducible representations (see Table \ref{tab0}),
three singlets $1,~1^\prime$ and $1^{\prime \prime}$ and one triplet.
The one-dimensional unitary representations are obtained by:
\begin{equation}
\begin{array}{lll}
1&S=1&T=1\\
1^\prime&S=1&T=\omega\\
1^{\prime\prime}&S=1&T=\omega^2
\end{array}
\end{equation}
where $\omega^3=1$. The product rule for the singlets are:
\begin{equation}
\begin{array}{l}
1\times1=1^\prime\times1^{\prime \prime}=1\\
1^{\prime}\times1^{\prime}=1^{\prime \prime}\\
1^{\prime \prime}\times1^{\prime \prime}=1^{\prime }
\end{array}\end{equation} 
In the basis
where $S$ is real diagonal,
\begin{equation}\label{eq:ST}
S=\left(
\begin{array}{ccc}
1&0&0\\
0&-1&0\\
0&0&-1\\
\end{array}
\right)\,;\quad
T=\left(
\begin{array}{ccc}
0&1&0\\
0&0&1\\
1&0&0\\
\end{array}
\right)\,;
\end{equation}
one has the following triplet multiplication rules, 
\begin{equation}\label{pr}
\begin{array}{lll}
(ab)_1&=&a_1b_1+a_2b_2+a_3b_3\,;\\
(ab)_{1'}&=&a_1b_1+\omega a_2b_2+\omega^2a_3b_3\,;\\
(ab)_{1''}&=&a_1b_1+\omega^2 a_2b_2+\omega a_3b_3\,;\\
(ab)_{3_1}&=&(a_2b_3,a_3b_1,a_1b_2)\,;\\
(ab)_{3_2}&=&(a_3b_2,a_1b_3,a_2b_1)\,,
\end{array}
\end{equation}
where $a=(a_1,a_2,a_3)$ and $b=(b_1,b_2,b_3)$. 

\section{$SU(5)$ embedding of the model}
\label{su5}

In \cite{Hirsch:2010ru} the quark sector has not been studied and it
is assumed that quarks are generically singlets of $A_4$ in order to
not couple to the DM.  It is possible to extend such a model to the
quarks by embedding it into the grand unified group $SU(5)$.  It is
beyond our scope to give a complete gran-unified model and we only
sketch a simple way to embed the model presented in
\cite{Hirsch:2010ru} into a grand unified group.  The matter
assignment is the following
%
\begin{center}
\begin{tabular}{ccccccccc}
\hline
&     $T_1$ &$T_2$ &$T_3$ &$F_1$ &$F_2$ &$F_3$ & $N_T$&$N_4$\\
\hline
SU(5)& 10& 10& 10& $\bar{5}$&$\bar{5}$&$\bar{5}$&1&1\\
\hline
$A_4$&1&$1'$&$1''$&1&$1''$&$1'$&3&1\\
\hline
\end{tabular}
\end{center}
where we have assumed three copies of ten-multiplets and three of
five-multiplets of $SU(5)$ to decribe the three flavours.  The scalar
assignment is
\begin{center}
\begin{tabular}{cccccccc}
\hline
SU(5)& $5_H$& $\bar{5}_H$& $5_\eta$& $45_H$\\
\hline
$A_4$& 1& $1$&$3$&1\\
\hline
\end{tabular}
\end{center}
then the Lagrangian is given by

\begin{eqnarray}
\mathcal{L}_{down}&=&
y_1^{l,d} T_1F_1\bar{5}_H+y_2^{l,d} T_2F_2\bar{5}_H+y_3^{l,d} T_3F_3\bar{5}_H
\quad +y_1^{\prime l,d} T_1F_145_H+y_2^{\prime l,d} T_2F_245_H+y_3^{\prime l,d} T_3F_345_H;\\
\mathcal{L}_{up}&=&
y^u_1 T_1 T_1 5_H+ y^u_2 T_2 T_3 5_H+y^{\prime u}_1 T_1 T_1 45_H+ y^{\prime u}_2 T_2 T_3 45_H;\\
\mathcal{L}_\nu&=&
y^\nu_1T_1 N_4 5_H+y^\nu_2T_2 N_4 5_H+y^\nu_3T_3 N_4 5_H+y^\nu_1T_1 N_T 5_\eta+M_1\,N_TN_T+M_2\,N_4N_4.
\end{eqnarray}

\vskip5.mm
The charged lepton and down quark mass matrix are diagonal with eigenvalues
\begin{equation}
\begin{array}{ccc}
m_e=y_1^{l,d}\langle 5_H \rangle-3y_1^{\prime l,d}\langle 45_H \rangle;&
m_\mu=y_2^{l,d}\langle 5_H \rangle-3y_2^{\prime l,d}\langle 45_H \rangle;&
m_\tau=y_3^{l,d}\langle 5_H \rangle-3y_3^{\prime l,d}\langle 45_H \rangle;\\
&&\\
m_d=y_1^{l,d}\langle 5_H \rangle+y_1^{\prime l,d}\langle 45_H \rangle;&
m_s=y_2^{l,d}\langle 5_H \rangle+y_2^{\prime l,d}\langle 45_H \rangle;&
m_b=y_3^{l,d}\langle 5_H \rangle+y_3^{\prime l,d}\langle 45_H \rangle;
\end{array}
\end{equation}
and the three charged lepton masses as well as the three down quark
masses can be easily reproduced.  The up quark mass matrix is
\begin{equation}
M^u=
\left(
\begin{array}{ccc}
m_u&0&0\\
0&0&m_c\\
0&m_t&0
\end{array}
\right),\quad
M^u\,M^{u\dagger}=
\left(
\begin{array}{ccc}
m_u^2&0&0\\
0&m_c^2&0\\
0&0&m_t^2
\end{array},
\right)
\end{equation}
where
\begin{equation}
\begin{array}{cccc}
m_u=y_1^u\langle 5_H \rangle;\quad
&
m_c=y_2^u\langle 5_H \rangle-y_2^{\prime u}\langle 45_H \rangle;\quad&
m_t=y_2^u\langle 5_H \rangle+y_2^{\prime u}\langle 45_H \rangle.&
\end{array}
\end{equation}
Given the structure of the up and down quark mass matrices, the quark
mixing matrix is diagonal. While this may be regarded as a good first
approximation, since quark mixing angles are small, clearly another
ingredient is needed such as, possibly, radiative corrections or extra
vectorlike quark states. A full fit of the quark sector observables within
a unified extension incorporating the flavour symmetry is beyond the
scope of our paper.

\section{Mass spectrum}
\label{spectrum}

To simplify the notation we define the following combinations of couplings:
\begin{eqnarray}
L&=&\lambda_9+\lambda_{10}+2\lambda_{11}, \\
Q&=&\lambda_{12}+\lambda_{13}+\lambda_{14}+\lambda_{15}, \\
P&=&\lambda_2 +\lambda_3 + 2\lambda_4+\lambda_5,  \\
R_1&=&(-3\lambda_3-6\lambda_4+2\lambda_6+\lambda_7+\lambda_8), \\ 
R_2&=&(-3\lambda_3-2\lambda_4-2\lambda_6+\lambda_7+\lambda_8-4\lambda_5), \\
R_3&=&(-3\lambda_3-4\lambda_4-2\lambda_5+\lambda_8).
\end{eqnarray}
The neutral scalars mass-matrix in the basis $H_0^{\prime}-H_1^{\prime}-A_0^{\prime}-A_1^{\prime}-H_2^{\prime}-H_3^{\prime}-A_2^{\prime}-A_3^{\prime}$
is block diagonal because of the $Z_2$ symmetry and CP-conservation. It is given by:

\begin{equation}\label{blockdiag}
M_{neutrals}=\left(
\begin{array}{cccc}
M_{HH_0}&0&0&0\\
0&M_{GA_0}&0&0\\
0&0&M_{H_2H_3}&0\\
0&0&0&M_{A_2A_3}\\
\end{array}
\right)\quad ,
\end{equation}

and the charged scalars mass matrix in the basis $H_0^{\prime+}-H_1^{\prime+}-H_2^{\prime+}-H_3^{\prime+}$ is:

\begin{eqnarray}\label{blockdiag}
M_{charged}=\left(
\begin{array}{cc}
M_{G^+H_0^+}&0\\
0&M_{H_2^+H_3^+}\\
\end{array}
\right).\quad 
\end{eqnarray}

The matrices $M_{GA_0}$ and $M_{G^+H_0^+}$ have a vanishing eigenvalue
corresponding respectively to the neutral and charged Goldstone bosons
eaten up by the $Z$ and $W^{\pm}$ gauge bosons.
The diagonalization of the mass matrices goes as follows:
\begin{eqnarray}
D_{HH_0}=U^{\dagger}_{HH_0}M_{HH_0}U_{HH_0};~~~~D_{GA_0}=U^{\dagger}_{GA_0}M_{GA_0}U_{GA_0};\nonumber \\
D_{G^+H_0^+}=U^{\dagger}_{G^+H_0^+}M_{G^+H_0^+}U_{G^+H_0^+};~~~~M_{X_2X_3}=U^{\dagger}_{23}M_{X_2X_3}U_{23},
\end{eqnarray}
with $X=H,A,H^+$ and
\begin{equation}
U_{GA_0}=U_{G^+H_0^+}=\left(
\begin{array}{ccc}
\cos\beta&-\sin\beta\\
\sin\beta&\cos\beta\\
\end{array}
\right)\,;\quad
U_{23}=\left(
\begin{array}{ccc}
-1/\sqrt{2}&1/\sqrt{2}\\
1/\sqrt{2}&1/\sqrt{2}\\
\end{array}
\right)\,;\quad
\end{equation}
The mixing angle between $\eta_2$ and $\eta_3$ is $\pi/4$.  The ratio
of the vevs was parametrized by the angle $\beta$ that controls the
mixing within the $Z_2$-even sector for the charged scalars and
pseudoscalars:
\begin{eqnarray}
  v^2={v_\eta}^2+{v_H}^2 ;\nonumber \\  
  \tan\beta=v_H/v_\eta.
\end{eqnarray}
We do not report the lenghty expression for $U_{HH_0}$.
The mass spectrum reads then as:
 \begin{eqnarray}
 \label{mass_eq}
  M_{A_0}^2&=&-2 v^2 \lambda_{11}, \\
  M_{H}^2&=&\lambda_1 {v_H}^2 +P {v_\eta}^2 -\sqrt({v_\eta}^2 {v_H}^2 (L^2-4 P \lambda_1)+(P {v_\eta}^2 +\lambda_1 {v_H}^2 )^2
  ), \\
  M^2_{H_0}&=&\lambda_1 {v_H}^2 +P {v_\eta}^2 +\sqrt({v_\eta}^2 {v_H}^2 (L^2-4 P \lambda_1)+(P {v_\eta}^2 +\lambda_1 {v_H}^2
  )^2 ),\\
  M^2_{H^+_{0}}&=&-(\lambda_{10}+2\lambda_{11}) v^2/2,\\
  M_{H_2}^2&=& v_\eta (-3Q v_H +R_1 v_\eta)/2,\\
  M_{A_2}^2&=&(-4\lambda_{11} {v_H}^2 + Q v_\eta v_H + R_2 {v_\eta}^2 )/2,\\
  M_{H^{+}_2}^2&=& ({v_H}^2 (\lambda_{10}+2\lambda_{11})-Q v_\eta v_H+R_3 {v_\eta}^2 )/2,\\
  M_{H_3}^2&=&{M^2_{H_2}}+3 Q v_H v_\eta, \\
  {M^2_{A_3}}&=&{M^2_{A_2}}+Q v_H v_\eta, \\
  {M^2_{H^{+}_3}}&=&{M^2_{H^{+}_2}}+Q v_H v_\eta.
\end{eqnarray}

\section{Oblique parameters}
\label{oblique}
Following the notation of \cite{Grimus:2007if}, the $T$ oblique parameter for the standard model extended by
$n$ higgs doublets with hypercharge $1/2$ is :

\begin{eqnarray}
\label{delat}
\Delta \rho &=& \alpha T= \frac{g^2}{64 \pi^2 m_W^2} \left\{
\sum_{a=2}^n\, \sum_{b=2}^{2n}\,
\left| \left( U^\dagger V \right)_{ab} \right|^2
F \left( m_a^2, \mu_b^2 \right)
\right. \label{1a} \\ & &
- \sum_{b=2}^{2n-1}\, \sum_{b^\prime = b+1}^{2n}\,
\left[ \mbox{Im} \left( V^\dagger V \right)_{b b^\prime} \right]^2
F \left( \mu_b^2, \mu_{b^\prime}^2 \right)
\label{1b} \\ & &
- 2\, \sum_{a=2}^{n-1}\, \sum_{a^\prime = a+1}^n\,
\left| \left( U^\dagger U \right)_{a a^\prime} \right|^2
F \left( m_a^2 , m_{a^\prime}^2 \right)
\label{1c} \\ & &
+ 3\, \sum_{b=2}^{2n}\,
\left[ \mbox{Im} \left( V^\dagger V \right)_{1b} \right]^2
\left[
F \left( m_Z^2, \mu_b^2 \right) - F \left( m_W^2, \mu_b^2 \right)
\right]
\label{1d} \\ & & \left.
- 3 \left[
F \left( m_Z^2, m_h^2 \right) - F \left( m_W^2, m_h^2 \right)
\right]
\right\},
\end{eqnarray}
where $m_a$, $m_a′$ denote the masses of the charged scalars and $\mu_b$, $\mu_b′$ are the masses
of the neutral ones, $\alpha$ is the fine-structure constant and the function $F$ is defined as ($x, y>0$):

\begin{equation}
\label{Eq:F}
F \left( x, y \right) \equiv
\left\{ \begin{array}{ll}
{\displaystyle \frac{x+y}{2} - \frac{xy}{x-y}\, \ln{\frac{x}{y}}}
&\Leftarrow\ x \neq y,
\\*[3mm]
0 &\Leftarrow\ x = y.
\end{array} \right.
\end{equation}
We evaluate the $U$ and $V$ matrices for our model as:

\begin{equation}
U=\left(
\begin{array}{ccc}
U_{G^+H_0^+}&0\\
0&U_{23}\\
\end{array}
\right)\,;\quad
V=\left(
\begin{array}{cccc}
i U_{G^+H_0^+}&U_{HH_0}&0&0\\
0&0&U_{23}&i U_{23}\\
\end{array}
\right).\quad
\end{equation}


\end{document}